\documentstyle[preprint,pre,aps,eqsecnum]{revtex}

\begin{document}

\draft
\preprint{IFUP-TH 11/95}
\draft
\title{The 2-dimensional non-linear $\sigma$-model on a
random lattice}
\author{B. All\'es and M. Beccaria}
\address{
Dipartimento di Fisica dell'Universit\`a  and INFN\\
Piazza Torricelli 2, I-56100 Pisa, Italy\\
}
\maketitle
\begin{abstract}
The $O(n)$ non-linear $\sigma$-model is simulated on 2-dimensional
regular and random lattices. We use two different levels of randomness
in the construction of the random lattices and give a detailed
explanation of the geometry of such lattices. In the simulations, we
calculate the mass gap for $n=3,\ 4$ and 8, analysing the asymptotic
scaling of the data and computing the ratio of Lambda parameters
$\Lambda_{\rm random}/\Lambda_{\rm regular}$. These ratios are in
agreement with previous semi-analytical calculations. We also
numerically calculate the topological susceptibility by using
the cooling method.
\end{abstract}

\vskip 5mm


\section{Introduction}

\vskip 5mm

Numerical simulations of an asymptotically free field theory on a
lattice
provide information about continuum physics when they are performed at
values of the correlation length which lay within the scaling
window. This region is defined by the inequalities $1 \ll \xi/a \ll L$
where $a$, $\xi$ and $L$ are the lattice spacing, the correlation
length and the lattice size respectively.

To this window it corresponds another scaling window
in terms of the bare coupling $g$. To understand the relationship
between both regions, it is enough to recall that
$\xi/a \propto \Lambda \exp(1/\alpha)$ where $\alpha$ is proportional
to some power of $g$. In this expression, $\Lambda$ is the
renormalization group independent mass parameter. This parameter
depends on the regularization used, thus the scaling window in terms of
the bare coupling can be shifted towards the region of lower couplings
if we use a lattice regularization in which $\Lambda$ is small. Gauge
theories regularized on a random lattice present a $\Lambda$ parameter
some orders of magnitude smaller than on a regular square
lattice~\cite{zhao}. Therefore, the simulations on a random lattice can
be performed at lower values of the bare coupling.

This fact may make the simulations on random lattices more
advantageous than on regular square lattices. The physical signal in
the Monte Carlo data can be masked by the presence of pertubative
expansions related to mixings, perturbative tails and finite
renormalizations of composite operators defined on the lattice.
If the expansion parameter is smaller (and the perturbative
coefficients do not get larger), these perturbative terms may be
better controlled and the non-perturbative physical signal be more
clearly seen. Also the non-universal terms in the scaling function
could be less relevant, thus expediting the asymptotic scaling.

The computation of the perturbative coefficients on random lattices
might seem rather involved. However, the higher degree of rotational
invariance on a random lattice should help in this
respect~\cite{inprogress}.

In order to address these prospects on a physically interesting
theory, like QCD, in reference~\cite{BA} one of us started to study
some features
concerning another asymptotically free field
theory, the $O(n)$ non-linear $\sigma$-model in two dimensions.
The action for this model in the continuum is
\begin{equation}
\label{action}
{\cal S} = {1 \over {2 g}} \int \hbox{d}^2x
(\partial_{\mu} {\vec \phi}(x))^2,
\end{equation}
\noindent with the constraint ${\vec \phi}(x)^2=1$
for all $x$.
In reference~\cite{BA} we chose this theory
because of its properties in common with QCD. Indeed, it is an
asymptotically free field theory~\cite{molti}
and presents a topological
content~\cite{belavin}
for $n=3$. Moreover, its mass gap is exactly known~\cite{maggiore}.
It was shown that
the first coefficients of the renormalization group functions
$\beta(g)$ and $\gamma(g)$ for this model regularized on a random
lattice are universal and that the $\Lambda$ parameter depends on the
degree of randomness $\kappa$ (see below)
used in the construction of the random lattice.

The present work has several aims.
Firstly, we want to verify the $\kappa$ dependence of
the $\Lambda$ parameter and support the scenario of a common continuum
limit, the same than that of the theory defined in Eq.~(\ref{action}).
For this purpose, the theory with $n=3$, 4 and 8 has been
simulated on both regular and random lattices to extract the
topological susceptibility and mass gap. We have also succesfully
used the cooling method~\cite{teper,control}
on a random lattice to extract the
topological content of the theory. The $n=3$ value was chosen to
study both the topological susceptibility and the mass gap, while
the higher values of $n$ were
used in order to have a better asymptotic scaling~\cite{plwolff}
on the mass gap data.
Another reason to study the topological charge on random lattices is
that some sites on these lattices can group together forming
clusters with a typical size less than one lattice spacing.
This geometry might allow the existence of very small
instantons which could hardly survive on regular lattices.
We have always used
lattices large compared to the correlation length at our $\beta$,
with $L=200$, 300 and 400. We have never averaged the
results of the simulations among different random lattices because the
previous lattice sizes are large enough to include an implicit average
among several subregions of the lattice. The updating has been
performed by using a cluster algorithm~\cite{wolff1}
adapted to the random lattice.

Another purpose of the present work is to analyse whether the
simulations on a random lattice improve the asymptotic scaling or not.
Actually, this is easy to check with the mass gap data
because the value of $m/\Lambda$ is known~\cite{maggiore}.
As it was shown in~\cite{BA},
$\Lambda_{\rm random} \approx \Lambda_{\rm regular}$ for
the $\sigma$-model. Hence, it is not expected any dramatic shift of
the scaling window towards small values of the bare coupling.

To construct the random lattice, we followed the procedure of
T. D. Lee at al.~\cite{td1}. The only new ingredient is the
introduction of a
degree of randomness $\kappa$. The sites of the lattice are the
centers of hard spheres, the radius of which is $a/2\kappa$. These
hard spheres are randomly located. At small values of $\kappa$, the
lattice is locally less random as we will see in section
2. There is a $\kappa_{\rm min}$ at which all distributions (link
lengths, distance between neighbouring sites, plaquette areas, etc.)
become Dirac deltas.

The plan of the paper is as follows. In section 2, we explain how to
construct a random lattice and discuss in detail both the $\kappa$
dependence and some geometrical properties like average link lengths
and average distance between neighbouring sites. We also give a
geometrical argument in favour of a unique random lattice, once
$\kappa$ is fixed, as the number of sites gets larger.
In section 3 we explain the simulation algorithm used. In section 4
we recall the scaling and $\beta$ functions for the non-linear
$\sigma$ model. We also explain the procedure followed to extract both
the mass gap and the topological susceptibility. A detailed
description of the cooling method is also given. In section 5
we show and analyse the Monte Carlo data for the physical
observables described in section 4. The conclusions are presented in
section 6.

\vskip 1cm

\section{Construction and geometry of a random lattice}

\vskip 5mm

We define a random lattice as a set of $N$ points located at
random on a volume $V$ with the condition that there are no two sites
closer than $a/\kappa$ where $a$ is the lattice spacing and $\kappa$
is a parameter to be fixed. These sites are placed with periodic
boundary conditions. Thus, we consider the 2-dimensional lattice as a
torus. In order to simulate a field theory on this
geometry, we must define a net of connections throughout the lattice,
linking neighbouring sites. In the first part of the section, we will
explain how to place the sites on the lattice and in the second part we
will review the triangularization process which is used to construct
the links and plaquettes. Finally, we will give some properties of the
random lattices.

If $N$ sites are placed on a 2-dimensional
volume $V$, we define the lattice spacing
$a$ as $a=\sqrt{V/N}$. The sites of our random lattice are the centers
of hard spheres (actually discs), the radii of which are $a/2\kappa$.
Thus, the sites are not closer than $a/\kappa$. Fixing the parameter
$\kappa \in [\kappa_{\rm min}, \infty]$
yields random lattices of different degrees of randomness as we will
see. The value of $\kappa_{\rm min}$ can be determined with the
following argument.
For large values of $\kappa$, the discs are small and they can be
placed loosely on the volume $V$. Then, the distributions
of link lengths, distances between sites, etc.\ display a wide
dispersion of values. Instead, for small $\kappa$ there is less
play in placing the discs and consequently the distributions
show a smaller  variance. On a regular square lattice, the hard
discs will
be placed as in Figure $1a$. In this case, $\kappa=1$. But also when
$\kappa=1$ there can be
a bit of dispersion in the distributions characterizing the
geometry of the lattice. For example, if we shift one column as
shown in Figure $1b$
\footnote{Due to the periodic boundary
conditions we imposed, the piece of disc which exceeds the allowed
volume from below, will reappear on the top of the lattice.}
then the distance $r_i$ from any site $i$ to its closest neighbour
still satisfies $r_i \geq a$. However,
the link lengths are no longer all equal to $a$. In particular, the
links joining the shifted sites with their non-shifted neighbours are
strictly longer than $a$.
Thus, a lattice constructed with $\kappa =1$ need not be regular.
The usual regular square lattice is only a particular case of the
lattice with $\kappa=1$.

 From these considerations, one can conclude that the smallest $\kappa$
is reached when the hard discs are packed tight as in Figure 2, forming
an hexagonal lattice. In this case, the hard nature of the discs
prevent the sites from shifting or moving from their positions in
the Figure. Therefore, all link lengths, distances between sites,
etc.\ are the same.
Stated in other words, the corresponding distributions
are Dirac deltas. As a consequence, the corresponding $\kappa$ must
be the smallest allowed.
The volume occupied by a regular hexagonal lattice with $N=L^2$ sites
is $V=L^2\sqrt{3}/2$, hence the parameter $\kappa$ is
$\kappa_{\rm min}= \sqrt{\sqrt{3}/2} \approx 0.9306049 \; $.

An explicit recipe for putting the $N$ sites on the volume of
the lattice is the following. The first site must be put at
random in any
position on the volume $V$. Now, imagine we have already placed
$\nu < N$ sites, then, by using the random generator, we propose
the coordinates of the $(\nu+1)$-th site and check
that its distance to all the $\nu$ previously
accepted sites is less than
$a/\kappa$. If the check is satisfied, we accept this site. If not, we
reject it and repeat the process by proposing another
$(\nu+1)$-th site.

This procedure is repeated until all $N$ sites have been placed
on the lattice. Clearly, the number of proposed sites $N_p$
satisfies $N_p \geq N$. In Figure 3 we plot the
experimental ratio $N_p/N$ as a function of $\kappa$
obtained during the construction of lattices with $100$ and $1000$
sites. From this plot, it is clear that the construction of random
lattices with small $\kappa$ is quite time consuming. One can
speed up the check about distances by
dividing the volume $V$ into boxes, large enough to contain a few
sites on average. Then, the check for a proposed site is performed in
the box it belongs to and in its neighbouring boxes.
However, even with this improvement, creating lattices
with $\kappa < 1.2$ is almost impossible with our recipe.
The smallest
value of $\kappa$ we have used in our simulations is $\kappa=1.3 \; $.

 The ratio $N_p/N$ depends only on $\kappa$, not on $N$.
Figure 3 confirms this statement. Let us derive a theoretical
expression for the curve in Figure 3 for large $\kappa$. If a site
cannot have neighbours closer than a distance $a/\kappa$, then once
$\nu$ sites have already been placed, the area left free for
putting new sites is
$V-\nu\pi (a/\kappa)^2$. This expression does not take into account
the fact that these
circles of radius $a/\kappa$ can overlap (with the constraint of
leaving their centers, the sites, not covered). For large $\kappa$,
the overlaps are less
frequent and the previous formula is adequate. Therefore, the
probability for a proposed point to be accepted after having put
$\nu$ sites is
\begin{equation}
   p_{\nu} = 1 - {{\nu \pi a^2} \over {V \kappa^2}}.
\end{equation}
\noindent Now, using $a^2=V/N$, we conclude that the total number
of proposed sites divided by $N$ is
\begin{equation}
   {1 \over N} \sum_{\nu=1}^N
     { 1 \over {1 - {{\nu \pi} \over {N \kappa^2}}}}.
\end{equation}
\noindent For large enough $N$, the sum becomes an integral giving the
ratio $N_p/N$ as a function of $\kappa$
\begin{equation}
 N_p/N \approx \int_0^1
 {{\hbox{d}x } \over { 1 - {{x\pi} \over \kappa^2}}} =
- {{\ln(1 - \pi/\kappa^2) \over {\pi / \kappa^2}}}.
\end{equation}
\noindent This expression, as previously stated, is independent of
$N$. It works well for $\kappa > 2.5 \; $.

Once the lattice volume $V$ has been filled with the $N$ sites for
some chosen value of $\kappa$, we proceed to the triangularization
process. We follow the method of reference~\cite{td2}.
It consists in joining sets of three sites to form a triangle with the
only condition that the circle circumscribed
by these three points does not
contain any other site. The sides of that triangle are the links
joining the three sites and the triangle itself is a plaquette.
This construction is unique and fills the whole lattice with
no overlapping among the triangles~\cite{td1}.
We also impose periodic boundary conditions in the triangularization
process.
In Figure 4 we show two $N=100$ sites random lattices,
with $\kappa=\infty$ and $\kappa=1.3 \; .$

It is useful to define also the dual lattice.
Its dual sites are the centers of the above-mentioned circumscribed
circles.
It is clear that any link is the common side of two triangles.
Thus, every link of the random lattice must be surrounded by
two dual sites. The line joining these two dual sites is called
the dual link.
Therefore, every link is associated with a dual link.

Let us call $l^{\mu}_{ij}$ the $\mu$-component of the link vector
that points from the site $i$ to the site $j$. The length of this
link is $l_{ij}=\parallel {{\bf l}_{ij}} \parallel$. The length of the
corresponding dual link is $s_{ij}$. Throughout this work, we will
often make use of the matrix $\lambda_{ij}$ defined as
\begin{equation}
\lambda_{ij} \equiv \cases
              {s_{ij}/l_{ij}, &if $i$ and $j$ are linked; \cr
               0, &otherwise. \cr}
\end{equation}
We introduce another vector, ${{\bf d}_{ij}}$ which is twice
the vector joining
the center of the vector link ${{\bf l}_{ij}}$ with the center of
the associated dual link ${{\bf s}_{ij}}$.
Another useful quantity is $\sigma_{ijk}$ defined as
\begin{equation}
 \sigma_{ijk} \equiv \left[ \left( {{\bf l}_{ij}} +
   {{\bf d}_{ij}} \right) \times \left( {{\bf l}_{ik}} +
   {{\bf d}_{ik}} \right) \right] \cdot {\hat {\bf z}}.
\end{equation}
\noindent In this
equation ${\hat {\bf z}}$ is the unit vector orthonormal
to the plane of the lattice, oriented as
${\hat {\bf z}} = {\hat {\bf x}} \times {\hat {\bf y}}$.

The set of dual links around the site $i$,
$\lbrace {{\bf s}_{ij}} \rbrace_{i \; {\rm fixed}}$,
determine a convex region called Voronoi cell. The area of this
cell is $\omega_i$.

As soon as the lattice has been constructed, one can devise tests
to check the triangularization. The first and easiest one is to
verify that the number of triangles (links) is equal to twice
(three times) the number of sites~\cite{td1}.
Other good tests are the integral
properties~\cite{td3}
\begin{equation}
\sum_j  \lambda_{ij} l^{\mu}_{ij} = 0,  \qquad \qquad \qquad
   \sum_{j} \lambda_{ij}
   l^{\mu}_{ij} (l^{\nu}_{ij} + d^{\nu}_{ij}) = 2 \omega_i
   \delta^{\mu \nu}.
\end{equation}
\noindent A third test consists in demanding that the
quadratic part of the
action has only one zero mode~\cite{BA}.

The action for the 2-dimensional
$O(n)$ non-linear $\sigma$-model on a random lattice can be written
as~\cite{td3}
\begin{equation}
 {\cal S}^L = {1 \over {4  g}}
\sum_{i,j} \lambda_{i j} ({\vec \phi}_i - {\vec \phi}_j)^2.
\end{equation}
\noindent In this equation, $g$ is the coupling constant.
$i, j, \dots$
denote sites and ${\vec \phi}_i$ stands for
the value of the field ${\vec \phi}$ at the site $i$.
This is the action we will make use in our numerical simulations,
both on regular and random lattices.
For a regular
square lattice, $\lambda_{ij}$ is 1 for linked sites
and zero otherwise.
Hence, Eq.~(2.7) becomes the standard action when it is
considered on a regular lattice.
One can show that the {\it na\"{\i}ve} continuum limit of Eq.~(2.7) is
the correct action for the model in the continuum, Eq.~(1.1).

 We see from Figure 4 that random lattices with
large (small) values of
$\kappa$ display a higher (lower) degree of randomness. As it was
shown in reference~\cite{BA},
random lattices with different $\kappa$ are
different regularizations of the same theory. In particular, it was
shown that the $\Lambda$ parameter is $\kappa$-dependent. In the
present paper, we will check this statement by a numerical simulation
and, moreover, we will give hints that the non-universal non-leading
coefficients of the $\beta$ function of the model are also
$\kappa$-dependent.

 The level of randomness can be clearly manifested with
the distributions of some geometrical properties of the lattice.
The first property we plot is the distance of one site to its nearest
site, $r$. This distance, referred to the site $i$, was written before
as $r_i$.
In Figure 5, we show the probability distribution of
this distance, $P(r) \hbox{d}r$ for a $\kappa=\infty$ lattice.
The histogram in the Figure is the numerical result, obtained by
calculating $r_i$ for a single site on 30000 random lattices
of 1000 sites. The
solid line is the theoretical well known Poisson distribution,
\begin{equation}
P(r) \hbox{d}r = 2 \pi {r \over {a^2}} \hbox{e}^{-\pi r^2 /a^2}
   \hbox{d}r.
\end{equation}
\noindent Both curves are normalized to 1 and should coincide.

 In Figure 6, we show the same distribution for a $\kappa=1.3$ random
lattice. It was obtained as in Figure 5, with 30000 random lattices of
1000 sites.
Notice that the distribution displays less dispersion. The
cutoff of the curve is placed at $a/\kappa$.
Given a random lattice constructed with some finite value of $\kappa$,
all sites are farther from each other than $a/\kappa$.
If we bring together two of these sites a distance less
than $a/\kappa$, the random lattice will not vary its geometrical
properties. We checked that the results of the simulations
performed on it, do not vary either.
Indeed, apart from a tiny bump at some position between $r=0$ and
$r=a/\kappa$, Figure 6 will not change. The more sites the random
lattice has, the tinier the bump becomes.

 Another quantity we can plot is the distribution of link lengths. In
Figure 7 we show the distribution of link lengths $P(l) \hbox{d}l$ for
a random lattice with $\kappa=\infty$. Given a site on the lattice, it
is always linked with the closest site to it. But it can also be
linked with sites placed quite far away.
This is why Figures 5 and 7 differ. Finally, in Figure 8 we show the
distribution of link lengths for a
random lattice with $\kappa=1.3 \; $.
Again we see that the distribution is sharper as $\kappa$ becomes
smaller.

In table 1 we show the numerical
average values of link lengths $\left\langle l \right\rangle$
and distance between closest sites
$\left\langle r \right\rangle$ for several values of $\kappa$.
These quantities can be used to characterize the degree of randomness.
However, our method to construct the random lattice needs only the
knowledge of $\kappa$ as previously explained. Therefore, we will
label our random lattices with this $\kappa$ parameter.
The column for $\left\langle l \right\rangle$
was obtained averaging the link lengths
of random lattices with 10000 sites for $\kappa=\infty, \; 1.3$
and 1000 sites for $\kappa= 1.2$.
The column for $\left\langle r \right\rangle$ was calculated by
averaging $r$ on a single site for 30000 random lattices of
1000 sites for $\kappa=\infty$ and $1.3$; for $\kappa=1.2$
we averaged on 14 random lattices of 100 sites.
The numbers shown depend on the lattice size. As this size
gets larger, the averages tend to stabilize. For instance, the
averages for $\left\langle r \right\rangle /a$
calculated on random lattices of 100, 1000
and 10000 sites with $\kappa=\infty$
are 0.523(2), 0.508(2) and 0.504(2) respectively.
The exact value computed from Eq.~(2.8) is
$\left\langle r \right\rangle = {1 \over 2}a$.
The errors shown are only statistical and do not include this
systematic effect.
The value of
$\left\langle l \right\rangle$
for $\kappa=\infty$ is known theoretically~\cite{td1},
it is $1.132 \, a$ in good agreement with the value shown in table 1.
The trend shown in Figures 5, 6 and 7, 8 is consistent with the fact
that as
$\kappa \longrightarrow \kappa_{\rm min}$ the distributions become
a Dirac delta. These Dirac deltas are equal and centered at
$l,r=\sqrt{2/\sqrt{3}} a \approx 1.0746 \, a$.

As the number of sites gets larger, all random lattices with a fixed
value of $\kappa$ become
equivalent. To support this statement we have generated 100 lattices
for several values of $L$ and verified that the average link length
presents less standard deviation among the 100 lattices
as $L$ becomes greater. If we define
$X \equiv D/\left\langle l \right\rangle$,
where $\left\langle l \right\rangle$
is the average link length among the 100 lattices for
a fixed value of $L$ and $D$ is the standard deviation of this
average, the results are $X=0.5045,\ 0.1994,\ 0.1116,\ 0.0807\ \%$ for
$L=20,\ 50,\ 100,\ 150$ respectively. This analysis was done with
$\kappa=\infty$ random lattices.

\vskip 1cm

\section{The simulation algorithm}

\vskip 5mm

The best algorithm for the updating of the $O(n)$ non-linear
$\sigma$-model in a numerical simulation is
the non-local Wolff algorithm~\cite{wolff1}. On the regular
lattice it does not show any critical
slowing down. Let $\tau$ be the time correlation of the Markov chain
of states generated
by the algorithm. Let $\xi$ be the spatial correlation length of the
model (a well defined
function of $\beta$). Then, the Wolff algorithm has the
fundamental property that
$\tau$ does not increase as the critical point, corresponding to
$\xi=\infty$, is approached.

This nice property is allowed by the intrinsic non locality of the
Wolff updating and has nothing to do with the details of the lattice. We
argue that it should hold also for the straightforward
generalization of the Wolff
algorithm on the random lattice. Indeed, our numerical data show
that the algorithm
performs very well at increasing $\beta$, just like in the regular case.

The above mentioned generalization is obtained as follows:
we start by choosing at random
one lattice site $i$ and an unitary vector $\vec{u}\in S^{n-1}$.
There is a great freedom in
the distribution of $\vec{u}$.
To be definite we chose the uniform distribution on the
hypersphere by taking at random
a point inside the hypercube $\vec{x}\in [0,1]^n$ and
rejecting it if $\parallel \vec{x} \parallel >1$.
Then, $\vec{u}=\vec{x}/\parallel \vec{x} \parallel$.

We now mark with a label the site $i$ and all of its neighbours
according to the probability weights
\begin{equation}
p_{ij} = 1-\exp\left[\min\left(0, -\frac{2\lambda_{ij}}{g}
(\vec{\phi}_i \cdot \vec{u})
(\vec{\phi}_j \cdot \vec{u})\right)\right].
\end{equation}
We continue recursively from the new marked
sites and proceed further until no
new site is added to the list of labelled sites.

When the recursive process ends, we collect
in a cluster all the marked sites and
transform them with a reflection
\begin{equation}
\vec{\phi} \to R\vec{\phi} \equiv \vec{\phi} - 2 (\vec{\phi} \cdot
\vec{u}) \vec{u}.
\end{equation}

As Wolff has shown~\cite{wolff2},
in the process of construction of the
cluster, all the quantities entering the r.h.s.\ of
\begin{equation}
\left\langle \vec{\phi}(x) \cdot \vec{\phi}(y) \right\rangle
= n\ N \left\langle
(\vec{\phi}(x)\cdot \vec{u})(\vec{\phi}(y)\cdot \vec{u})
\frac{1}{|C|} \theta_C(x)\theta_C(y) \right\rangle
\end{equation}
have already been calculated at each updating step. The above
expression is called an improved estimator for the two point function.
In Eq.~(3.3) $N$ is the total number of sites of the lattice,
$|C|$ is the number of sites in the cluster and
$\theta_C$ is the characteristic function of the cluster.
Another advantage of this estimator is that it can be measured after
each updating step without need of decorrelating updatings.

\vskip 1cm

\section{The physical observables}

\vskip 5mm

The physical observables that we have measured are the mass gap
for several values of $n$
and the topological charge for $n=3$.
For each of these quantities we have studied the asymptotic
scaling behaviour.
The universal 2-loop $\beta$ function of the $O(n)$ non-linear
$\sigma$-model is
\begin{equation}
\beta(g) = -\frac{n-2}{2\pi}g^2 - \frac{n-2}{4\pi^2}g^3 + O(g^4),
\end{equation}
and therefore the lattice spacing obeys the renormalization group law
\begin{equation}
a \Lambda_{\rm lattice} = f(\beta) \equiv
\left(\frac{2\pi\beta}{n-2}\right)^{1/(n-2)}
\exp\left(-\frac{2\pi\beta}{n-2}\right)
\left(1 + {\cal O}(\frac{1}{\beta})\right).
\end{equation}
The first non-universal correction for this scaling function on a
regular
lattice is known~\cite{falcionitreves}. On the other
hand, the value of $\Lambda_{\rm lattice}/\Lambda_{\overline{\rm MS}}$
is known both for regular~\cite{buh} and random lattices~\cite{BA}.

Any dimensionful quantity ${\cal A}$ measured on the lattice with
the operator ${\cal A}^L$ must satisfy (after the subtraction of
the relevant mixings)
\begin{equation}
\frac{{\cal A}^L}{f(\beta)^{\dim {\cal A}}} =
\frac{{\cal A}}{\Lambda_{\rm lattice}^{\dim {\cal A}}},
\end{equation}
when $\beta\to\infty$. Asymptotic scaling holds when the l.h.s.\ of
the above expression,
with the first terms of the beta function in $f(\beta)$,
depends little on $\beta$.

The analytical prediction for the mass gap of the $\sigma$ model
is~\cite{maggiore}
\begin{equation}
m = \left(\frac{8}{e}\right)^{\Delta}\frac{1}{\Gamma(1+\Delta)}\
\Lambda_{\overline{\rm MS}},
\qquad\qquad \Delta = \frac{1}{n-2},
\end{equation}
which combined with the ratio~\cite{buh}
\begin{equation}
\Lambda_{\rm regular} =
32^{-1/2}\exp\left(-\frac{\pi \Delta}{2}\right)\
\Lambda_{\overline{\rm MS}},
\end{equation}
gives $m/\Lambda_{\rm regular}$.

In order to evaluate $m$ we studied the wall-wall correlation function
by integrating the two-point function along the spatial direction and
studying it at
large temporal separations. Notice that on the random lattice the
$\nu$-th temporal slice
must be defined as the set of sites with
$t_{\nu} < t < t_{\nu+1}$.
Moreover, the number of sites that this time slice contains is a
function of $\nu$. This is a source of systematic error.
As $N$, the total number of sites, gets larger, the
width of the time slice $t_{\nu+1}-t_{\nu}$ can be smaller
and reduce the previous systematic error.

Numerical studies on the regular lattice indicate that
a fit of the wall-wall correlation Monte Carlo data
to the behaviour
\begin{equation}
C(t) \equiv A\ \cosh\left[B\left(t-\frac{L}{2}\right)\right]
\end{equation}
is enough to reproduce the data~\cite{wolff2}.
To avoid correlation among data at different $t$, we used sets of
different runs for every $t$. Hence, the statistical errors
obtained are reliable.

The above mentioned improved estimator, Eq.~(3.3),
reduces greatly the statistical noise
which affects this measure particularly at large $t$.

We have also measured the topological charge and susceptibility of the
$O(3)$ model on a random lattice. The topology of this model is based
on the stereographic map from the sphere $S^2$ onto the projective
plane~\cite{belavin}.
The topological charge actually counts the winding or instanton
number of this mapping on classical configurations. In the continuum
it is defined as
\begin{equation}
Q = \int \hbox{d}^2x Q(x) , \qquad
Q(x) = \frac{1}{8\pi}\epsilon_{\mu\nu}
\epsilon_{abc} \phi^a(x)\partial_\mu\phi^b(x)\partial_\nu\phi^c(x),
\end{equation}
and it rigorously yields integer values on a smooth field configuration.
At the quantum level, $Q$ is a composite operator which requires a
renormalization procedure.

On a random lattice, we introduce the following definition of
topological charge at the site $i$
\begin{equation}
Q_i = \frac{1}{32\pi}\
\sum_{j,k}\frac{\lambda_{ij}\lambda_{ik}}{\omega_i}\
\epsilon_{abc}\ \phi_i^a\phi_j^b\phi_k^c\ \sigma_{ijk}
\end{equation}
which, by using Eqs. (2.5) and (2.6), reproduces the continuum
expression, Eq.~(4.7). On a regular lattice, where
${\bf d}_{ij}={\bf 0}$, ${\bf l}_{ij}$ is equal either to
${\hat {\bf x}}$
or to ${\hat {\bf y}}$ and $\lambda_{ij}$ is 1 for linked sites and 0
otherwise, Eq.~(4.8) reproduces the standard
definition~\cite{vicarietal}.
The total topological charge $Q$ is calculated just by $Q=\sum_i Q_i$.

An interesting and non-trivial quantity which can be extracted from the
topological charge is the topological
susceptibility which in the continuum is defined as the two point
correlation of the topological charge at zero external momentum
\begin{equation}
\chi \equiv \int \hbox{d}^2x \left\langle Q(x) Q(0) \right\rangle.
\end{equation}
On the lattice, this is rewritten in the following way
\begin{equation}
\chi^L = \frac{1}{V}\left\langle\left(\sum_iQ_i\right)^2 \right\rangle.
\end{equation}

The Monte Carlo data for the topological susceptibility $\chi^L$
is~\cite{vicarietal}
\begin{equation}
\chi^L = a^2(\beta) Z(\beta)^2 \chi + a^2(\beta)
A(\beta){\left\langle S(x) \right\rangle}_{NP} + P(\beta),
\end{equation}
where $a(\beta)=f(\beta)/\Lambda_{\rm lattice}$, Eq.~(4.2),
${\left\langle S(x) \right\rangle}_{NP}$
is the non-perturbative
vacuum expectation value of the density of action
and $Z(\beta)$, $A(\beta)$ and $P(\beta)$ are the renormalizations
which can be calculated perturbatively~\cite{vicarietal,tutti}.
On the regular lattice
${\left\langle S(x) \right\rangle}_{NP}$
is negligible~\cite{papa} and so is
the second term in the r.h.s.\ of Eq.~(4.11). We have not calculated
all of these renormalizations on the random lattice but instead we
have used the cooling method~\cite{teper}.
The cooling procedure is a relaxation
process which after a few cooling steps ($\sim$ 30--40 steps) have
almost eliminated all short scale fluctuations leaving the long waves
still present. If we assume that these short scale fluctuations, of
order ${\cal O}(a)$, are responsible for the quantum noise which shows
up as renormalization effects, then after some cooling steps all
renormalizations in Eq.~(4.11) will disappear and the
physical and Monte Carlo topological susceptibility will be related by
the expression
\begin{equation}
\chi = \frac{\chi^L}{a(\beta)^2}.
\end{equation}
However the situation for the $O(3)$ $\sigma$ model is not so simple.
In this model, instantons tend to be small. Indeed the
distribution of instantons with radius $\rho$
in this theory satisfies
$\hbox{d}{\cal N}/\hbox{d}\rho \propto 1/\rho$.
As a consequence,
the previous cooling process can also eliminate small ${\cal O}(a)$
instantons, thus modifying the topological content of the
configuration. This unwanted behaviour of the cooling occurs
on regular lattices~\cite{papa}.
On a random lattice some sites can group together forming
clusters with a typical size less than one lattice spacing.
This fact happens mostly on large $\kappa$ lattices (see Figure~4).
This geometry might allow the existence of very small
instantons with a size $\rho \ll a$. This is the main motivation for
studying the topological properties of the model on random lattices.

In Figure 9 we show the evolution of the topological charge for 40
uncorrelated configurations as the cooling process goes on. At zero
cooling step the value of the charge on the lattice is on average
$Q^L=Q Z(\beta)$. Instead, after 30 or 40 steps, this charge
reached an almost integer value which depends only on the underlying
instantonic content of the original configuration.
This almost integer value remains stable for a long plateau.
Thus, we assumed the value of $Q^L$ after 30 or 40 steps of cooling
as the correct topological charge of the configuration.
We checked that the value of the topological susceptibility is also
stable on the plateau. Indeed, the value obtained is within
errors, the same if 30, 40 or 50 cooling steps are performed.
It can also be seen from Figure 9 that the value of $Q^L$ after
several coolings is not exactly an integer. This is a general fact and
has to do with the fact that instantons are not exact solutions of the
theory defined on a lattice. We also checked that the susceptibility
is insensitive to rounding this number to its nearest integer.
We chose for our analysis the values of $Q^L$ after 30 cooling steps
without rounding it to the nearest integer.

We now turn to a detailed description of the cooling step.
It consists in locally minimizing the action with respect to the field
at each site once per step.
We used a controlled cooling~\cite{control}.
This means that for a given positive
number $\delta$, the new field
$\vec{\phi}_i'$ and the old one $\vec{\phi}_i$ differ less than
$\delta$, $\parallel \vec{\phi}_i - \vec{\phi}_i' \parallel \leq
\delta$. First we define the force relative to the site $i$ as
\begin{equation}
 \vec{F}_i \equiv \frac{\sum_j \lambda_{ij} \vec{\phi}_j}
{\parallel \sum_j \lambda_{ij} \vec{\phi}_j \parallel},
\end{equation}
and then we compute the distance between the force and the field at
this site $i$, $d \equiv \parallel \vec{F}_i - \vec{\phi}_i
\parallel$. Now, if $d \leq \delta$ then the new field is going to be
$\vec{\phi}_i' = \vec{F}_i$. Instead, if $d > \delta$ then the
field at the site $i$ becomes
\begin{equation}
\vec{\phi}_i' = \frac{\vec{\phi}_i +
\epsilon\ (\vec{F}_i - \vec{\phi}_i)}
{\parallel \vec{\phi}_i + \epsilon\
(\vec{F}_i - \vec{\phi}_i) \parallel},
\end{equation}
where $\epsilon$ is chosen to be
\begin{equation}
\epsilon \equiv
\frac{\delta}{d \sqrt{1 - {1 \over 4} d^2}}.
\end{equation}
In a step of cooling we pass through all sites $i$ of the lattice
and perform the
previous modification on the corresponding field $\vec{\phi}_i$.

\vskip 1cm

\section{The Monte Carlo results}

\vskip 5mm

In this section we will show the set of Monte Carlo data obtained for
the mass gap and the topological susceptibility and discuss their
consequences.

In table 2 we show the set of data for the mass gap and the $O(3)$
$\sigma$ model on lattices of
$200^2$ sites. They are also shown in Figure 10.
The data were obtained from 12000 measures of the improved correlator,
Eq.~(3.3) for each $\beta$.
We fit these data to
the scaling function of Eq.~(4.2)
\begin{equation}
f(\beta) = 2 \pi \beta \alpha_1 \exp(-2 \pi \beta \alpha_2),
\end{equation}
where $\alpha_1$ and $\alpha_2$ are the parameters of the fit. In
particular, $\alpha_2$ must be equal to $1$
in order for the data to scale
as the $\beta$ function of the theory predicts, Eq.~(4.1), and
$\alpha_1$ must be $m/\Lambda_{\rm lattice}$.

The fit gives $\alpha_2=1.02(2),\ 0.97(2),\ 0.95(2)$ for
the regular lattice, $\kappa=1.3$ and $\kappa=\infty$ random lattices
respectively. This is in agreement with the result of~\cite{BA}
concerning
the universality of the random lattice: the first coefficient of the
beta function is the same in any lattice regularization.
Now, imposing $\alpha_2=1$ and leaving $\alpha_1$ free we obtained
$\alpha_1=121(2),\ 139(2),\ 91(1)$ for regular lattice, $\kappa=1.3$
and $\kappa=\infty$ random lattice respectively, with a value for
$\chi^2/{\rm n.d.f.}$ equal to 6/11, 8/11 and 9/12 respectively.
Following~\cite{maggiore}
and~\cite{BA}, the expected results for $\alpha_1$ are 80.09, 87.05 and
44.49 respectively. We excluded any finite size explanation to this
discrepancy. We arrived at this conclusion because $a)$
the technique of reference~\cite{luscherbender}
did not improve the results and $b)$ the
fits with the data of table 3 for $400^2$ lattices yielded similar
results for $\alpha_1$.
We think that the disagreement is due to the fact that our
range of $\beta$ is narrow enough to collect into $\alpha_1$
all the power-law corrections to the asymptotic scaling function.
Unfortunately, another single free parameter is not able to account for
the whole non-universal terms correcting Eq.~(5.1) and it did not
improve dramatically the result for $\alpha_1$.

 From the results of the fits for $\alpha_1$ we can get the ratio
between $\Lambda$ parameters.
Let us define $R(\kappa) = \Lambda_{\rm random}/\Lambda_{\rm regular}$
at a given $\kappa$. Then,
$R(1.3)=0.87(3)$ and $R(\infty)=1.33(4)$.
These results are in agreement with the average of the ratios obtained
from each $\beta$: $R(1.3)=0.86(6)$ and $R(\infty)=1.29(8)$.
These ratios for each $\beta$ are shown in Figure 11.
The theoretical values~\cite{BA} are
$R(1.3)=0.92(2)$ and $R(\infty)=1.8(2)$.
For $\kappa=1.3$ theoretical and Monte Carlo ratios are in agreement
within errors. But this is no so for $\kappa=\infty$. We again think
that this is due to the lack of asymptotic scaling.

It is well known that
asymptotic scaling is rather elusive in the $O(3)$ $\sigma$ model and
here we have realised that this problem
does not ameliorate if a random
lattice is used. For this reason, we also performed runs for the
$O(4)$ and $O(8)$ models where it is known that asymptotic scaling is
better achieved~\cite{plwolff}.
In table 4 the mass gap data for $O(4)$ on lattices
with $400^2$ sites are shown. In table 5 the same data are shown for
$O(8)$ and lattices with $300^2$ sites. In both cases 12000 measures
of the improved estimator, Eq.~(3.3), were performed for each $\beta$.

The $O(4)$ data of table 4 were fitted to
\begin{equation}
f(\beta)=\sqrt{\pi \beta} \alpha_1 \exp(-\pi \beta \alpha_2).
\end{equation}
The result for $\alpha_2$ was again close to 1: 1.02(4), 1.05(7) and
1.04(4) for regular, $\kappa=1.3$ and $\kappa=\infty$ random lattices
respectively. Fixing $\alpha_2$ to 1, we obtained for $\alpha_1$
26.6(5), 28.4(9) and 22.3(4) for for regular, $\kappa=1.3$ and
$\kappa=\infty$ random lattices respectively, while the predicted
values are~\cite{BA,maggiore} 24.02, 25.55 and 16.01$\;$.
We see a much better
agreement between theoretical and Monte Carlo results. The ratio of
$\Lambda$ parameters extracted from the Monte Carlo data is
$R(1.3)=0.94(5)$ and $R(\infty)=1.20(4)$.
The theoretical ratios are~\cite{BA} 0.94(2) and 1.5(1) respectively.
The agreement for $\kappa=\infty$ is still not satisfactory.

The data of table 5 for $O(8)$ were fitted to
\begin{equation}
f(\beta)=\left( \frac{\pi \beta}{3} \right)^{1/6} \alpha_1
\exp\left(- {{\pi \beta \alpha_2} \over 3} \right).
\end{equation}
Again $\alpha_2$ was satisfactorily close to 1. The fit for $\alpha_1$
shows a good asymptotic scaling ($\alpha_1=10.0(4)$
for the regular lattice while Eqs. (4.4)
and (4.5) give 9.48). So we expect that this time the ratios of
$\Lambda$ parameters extracted from Monte Carlo will be in agreement
with the theoretical ones. The Monte Carlo ratios are
$R(1.3)=0.90(6)$ and
$R(\infty)=1.29(9)$.
The theoretical ratios are 0.95(2) and 1.3(1) respectively.

We conclude that the dependence of the $\Lambda$ parameter on $\kappa$,
the degree of randomness of the lattice, is qualitatively correct.
The figures shown in~\cite{BA}
for these ratios are also correct but the
lack of asymptotic scaling prevent us from checking them for the
$O(n)$ model when $n \leq$ 6--8.

Let us now discuss the results for the topological susceptibility
obtained for the $O(3)$ model. In table 6 we show $\chi^L$ after 30
cooling steps which, as described in section 4, equals $a^2 \chi$.
These data have been obtained on lattices with $200^2$ sites
performing the cooling process on 1000 uncorrelated configurations
for each $\beta$.
In Figure 12
we show this set of data for the three kinds of lattices
used and the result of the fits performed on it by imposing the
scaling function Eq.~(5.1) with $\alpha_2=1$. It is apparent that the
data do not scale as Eq.~(5.1) imposes.

If $\alpha_2$ is left free, we see again
that the data do not scale as they
should. Indeed, for the regular lattice, $\kappa=1.3$ and
$\kappa=\infty$ random lattices, we obtain $\alpha_2=0.70(4),\
0.64(4),\ 0.69(4)$ respectively. We repeated the same analysis on
larger lattices to check whether this discrepancy is due
to finite size effects on the data.
In table 7 we show the data
for a regular lattice with $300^2$ sites and in table~8 the data for a
$\kappa=\infty$ random lattice with $400^2$ sites.
They are also obtained by cooling 1000 uncorrelated configurations.
The fit on the data
of table 7 gives $\alpha_2=0.76(4)$ and the fit on table~8
$\alpha_2=0.65(4)$. From these figures we could hardly conclude that
the lack of correct scaling is due to finite size effects.
We think that this problem must be traced back to the elimination of
small instantons during the cooling process. Probably, the use of
large $\kappa$ random lattices is not enough to avoid this effect.

The lack of asymptotic scaling
prevented us from using the topological susceptibility and the mass
gap data to check the physical
scaling of the $O(3)$ model on random lattices.

In Figure 13
we show the ratio of $\Lambda$ parameters for each value of
$\beta$ from the data of the topological susceptibility. The average
result for this ratio is surprisingly similar to the one obtained from
the mass gap data:
$R(1.3)=0.94(2)$ and
$R(\infty)=1.37(3)$.

\vskip 1cm

\section{Conclusions}

\vskip 5mm

We have used the random lattice to simulate the 2-dimensional
$O(n)$ non-linear $\sigma$-model. The sites of the random lattice are
considered as the centers of hard spheres of radius $a/2\kappa$ where
$a$ is the lattice spacing. These hard spheres are located at
random on the lattice volume.
The links between neighbouring sites are
established by a well known triangularization process~\cite{td2}.

To compare the performance of different lattices, we made the
simulations on regular lattices and random lattices with both
$\kappa=1.3$ and $\kappa=\infty$.

We used the Wolff algorithm~\cite{wolff1} for
the simulations as well as an
improved estimator~\cite{wolff2} for
the computation of two-point correlation
functions. We measured the mass gap (as the inverse of the correlation
length measured from the wall-wall two-point correlation function) and
the topological susceptibility. The topological charge was calculated
by using a cooling technique~\cite{teper} and we introduced a regularized
operator
for the topological charge on the lattice (see Eq.~(4.8)).

The Monte Carlo results for the mass gap scale as they should,
confirming previous claims~\cite{BA}
about the universality of the random
lattice regularization. They do not present any finite size problems
(for $200^2$ sites) but for small values of $n$ the asymptotic scaling
is not fulfilled. For the $O(8)$ model, where the data display a good
asymptotic scaling, we can reproduce the theoretical value for the
ratio between Lambda parameters
$\Lambda_{\rm random}/\Lambda_{\rm regular}$ for $\kappa=\infty$,
thus confirming the
semi-analytical prediction of reference~\cite{BA}.
Instead, for $\kappa=1.3$, the Monte Carlo value for the previous
ratio is in good agreement with the theoretical prediction already in
the $O(3)$ model. This could mean that both regularizations (regular
lattice and $\kappa=1.3$ random lattice) are quite similar and in
particular the non-universal terms in the scaling function, Eq.~(4.2),
almost coincide.
This assumption is also supported by the fact that
$\Lambda_{\rm random}/\Lambda_{\rm regular}$ for $\kappa=1.3$ is close
to 1. In any case, these conclusions are consistent with the scenario
where also the non-universal terms in the scaling function, Eq.~(4.2)
are $\kappa$-dependent.

The data for the topological susceptibility scale very badly.
We think that the cooling process removes small instantons thus
modifying the topological content of the configuration also on random
lattices.
The cooling smooths out fluctuations with a length of order
${\cal O}(a)$.
For smaller $\beta$ the lattice spacing is longer in physical units,
therefore the
number of eliminated instantons when smoothing out
${\cal O}(a)$ fluctuations is also larger. This explains why the data
in Figure 12 and Tables 6,7 and 8 are shifted
downwards for small $\beta$.

In any case, our results prove that the semi-analytical method used in
reference~\cite{BA}
is reliable to perform analytical calculations on
random lattices.

\vskip 1cm

\section{Acknowledgements}

\vskip 5mm

We thank Federico Farchioni and Andrea Pelissetto for useful
discussions. We also acknowledge financial support from INFN.


\newpage

\noindent{\bf Figure captions}

\begin{enumerate}

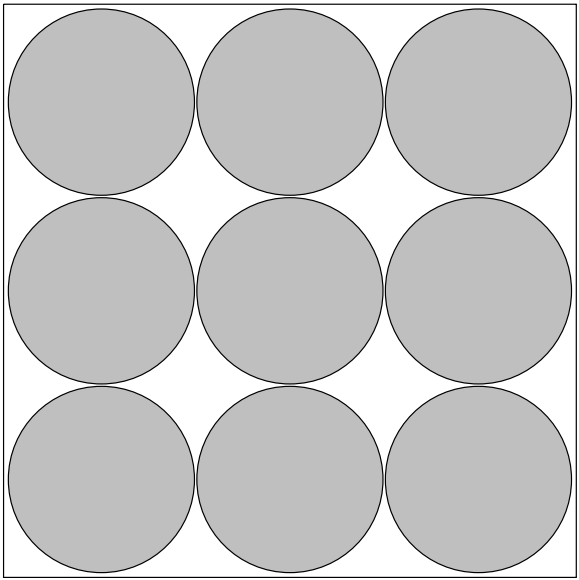
\item[Figure 1.] Location of hard discs for a lattice of $3^2$ sites
with $\kappa =1$. The radius of the discs is equal to $a/2$.
In a) it corresponds to a regular square lattice.
In b) a column has been shifted: $\kappa$ still equals 1 but the
lattice is no longer regular.

\vskip 3mm

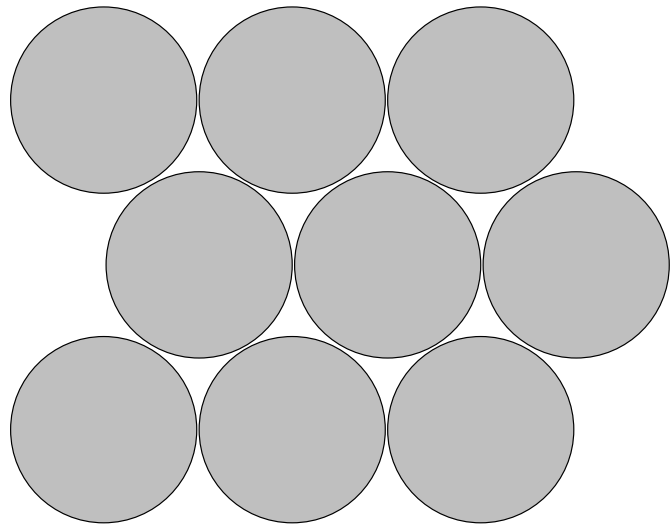
\item[Figure 2.] Location of hard discs for a regular hexagonal
lattice with 9 sites. They fit tight and no movement is allowed
without breaking the hard nature of the discs.

\vskip 3mm

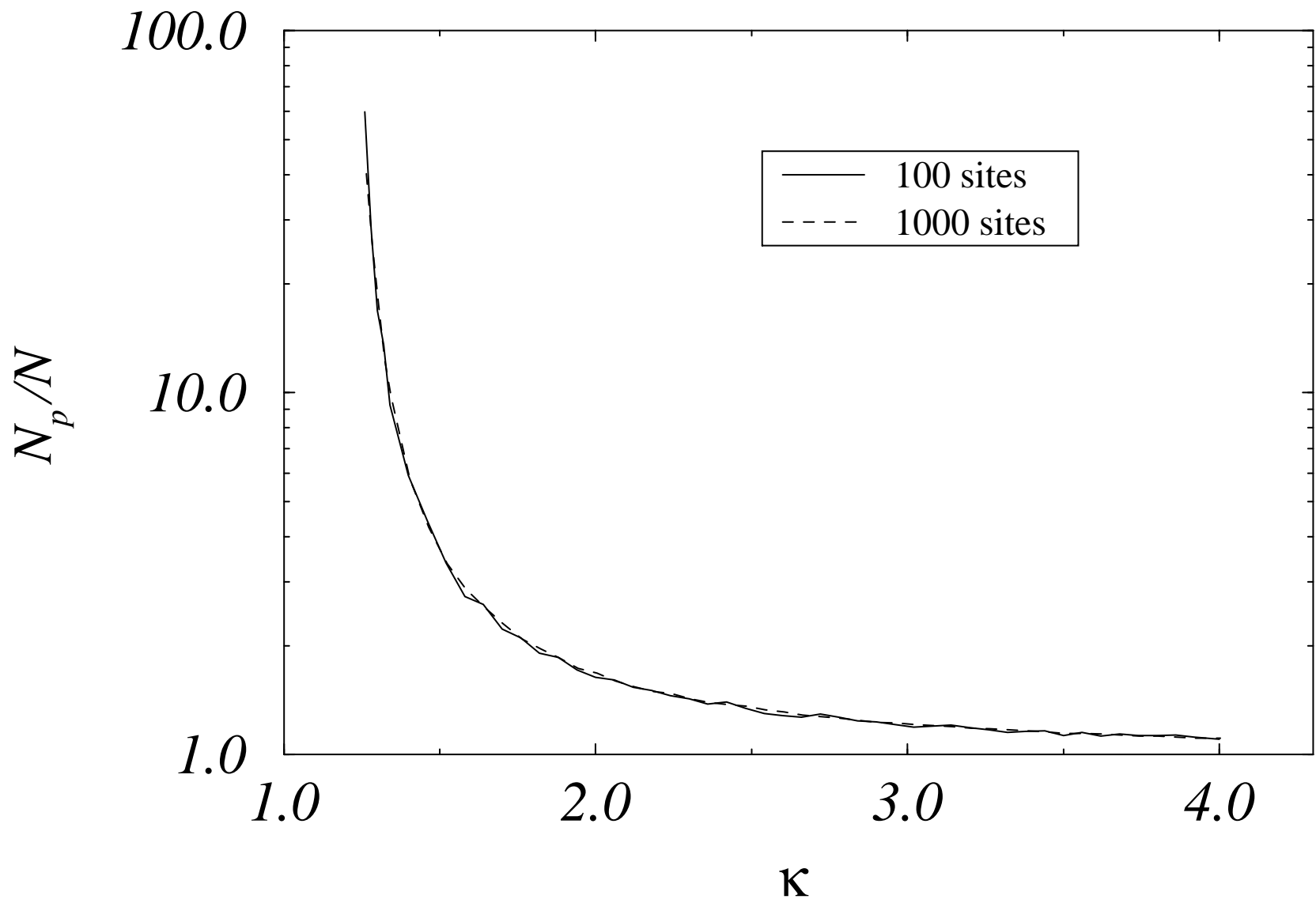
\item[Figure 3.] The ratio between the number of proposed sites $N_p$
and total number of sites $N$ in the construction of a random lattice
as a function of $\kappa$. The solid and dashed lines are the results
for lattices with $N=100$ and $N=1000$ sites respectively.

\vskip 3mm

\item[Figure 4.] The result of the triangularization process performed
on two random lattices of $10^2$ sites for a) $\kappa=\infty$ and b)
$\kappa =1.3 \;$. The different level of randomness is apparent.

\vskip 3mm

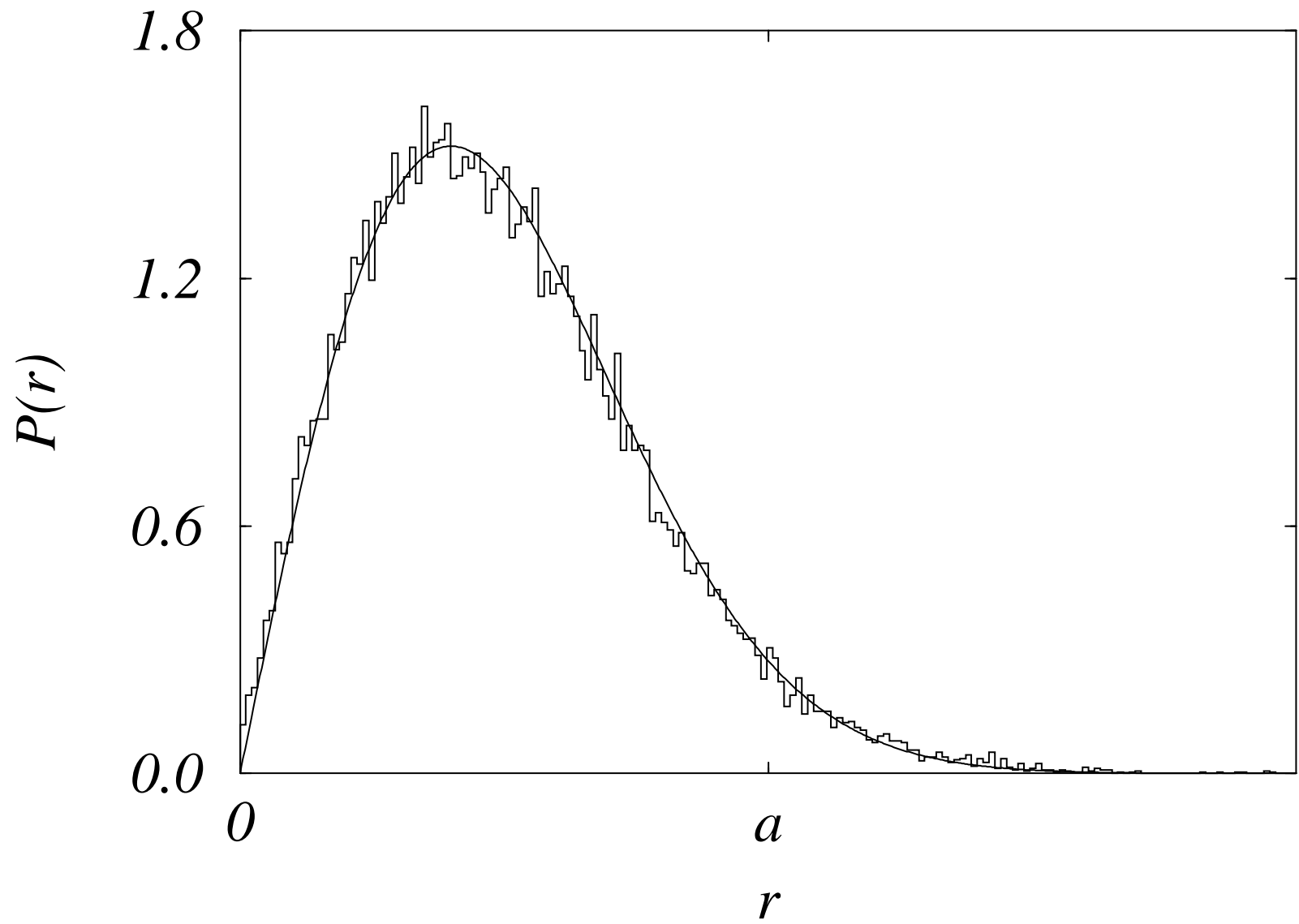
\item[Figure 5.] Distribution probability of distances between a site
and its closest neighbour on a random lattice with $\kappa=\infty$.
The $a$ in abscisses stands for one lattice spacing. The
histogram is the numerical result calculated by using a single site on
30000 random lattices of 1000 sites. This histogram should coincide
with the Poisson distribution, shown in the figure with a solid line.
Both curves are normalized to 1.

\vskip 3mm

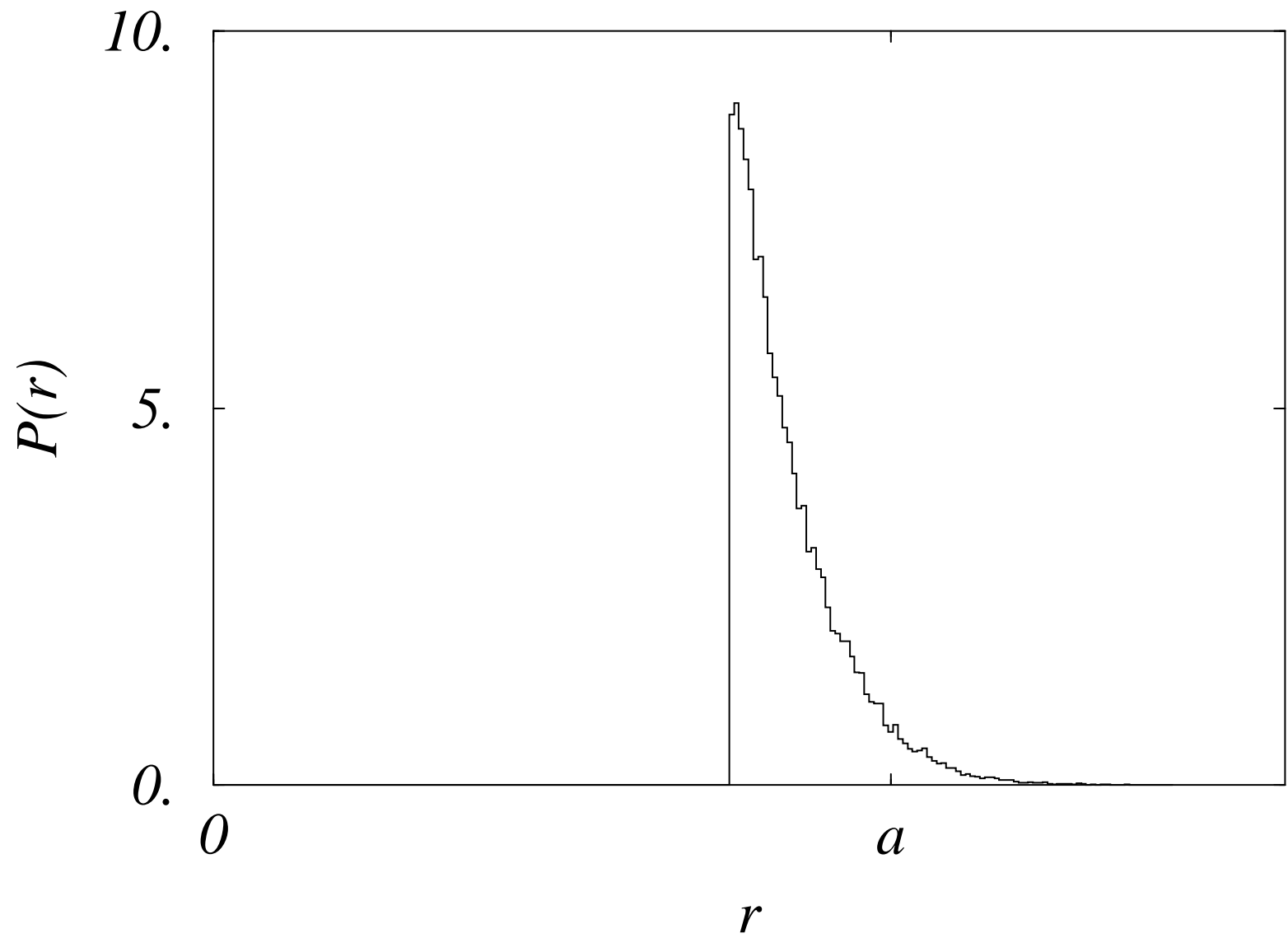
\item[Figure 6.] Distribution probability of distances between a site
and its closest neighbour on a random lattice with $\kappa=1.3 \;$.
The $a$ in abscisses stands for one lattice spacing. The
histogram is the numerical result calculated by using a single site on
30000 random lattices of 1000 sites. The curve is normalized to 1.

\vskip 3mm

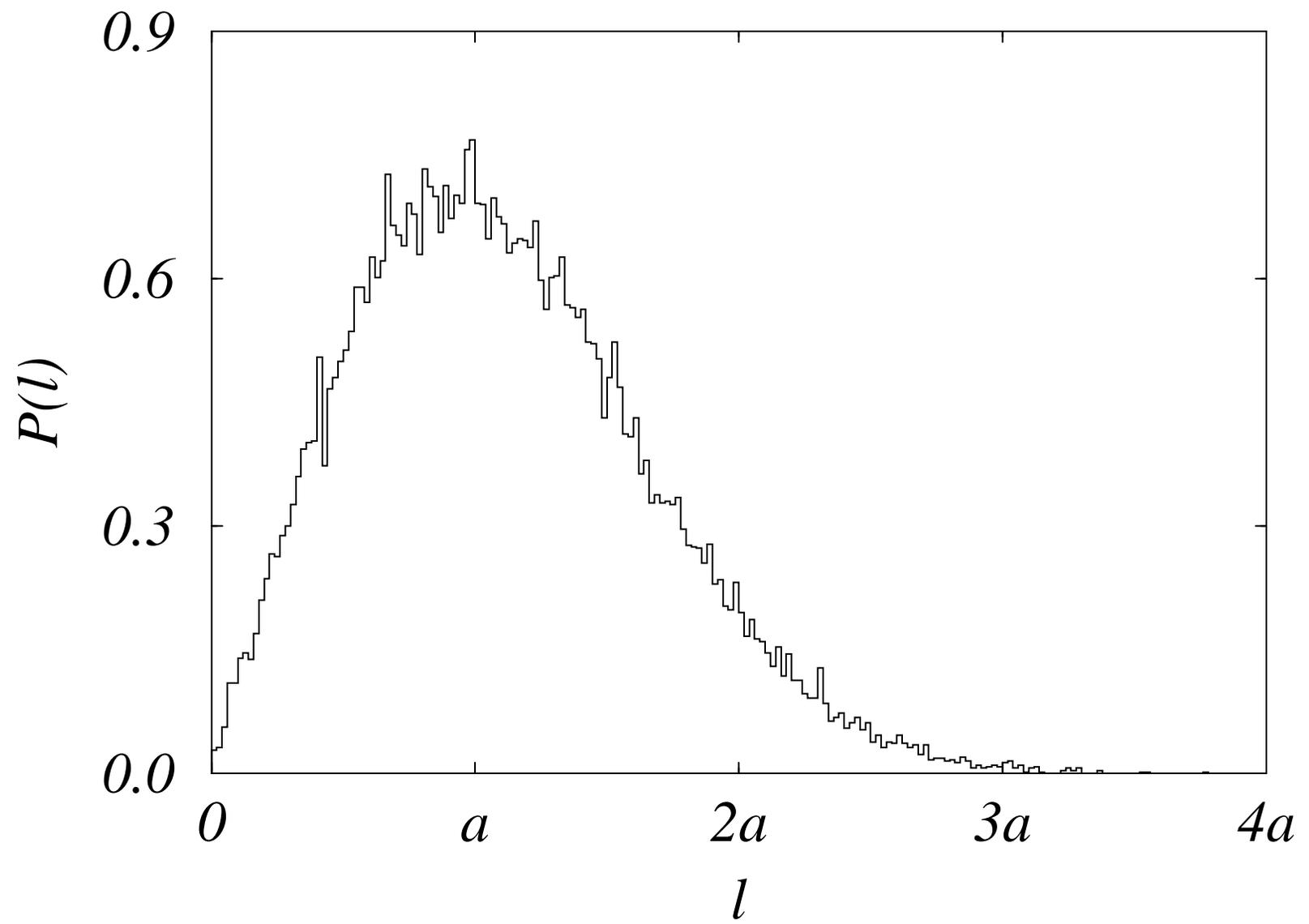
\item[Figure 7.] Distribution probability of link lengths on a random
lattice calculated with $\kappa=\infty$.
The $a$ in abscisses stands for one lattice spacing.
The curve is normalized to 1.

\vskip 3mm

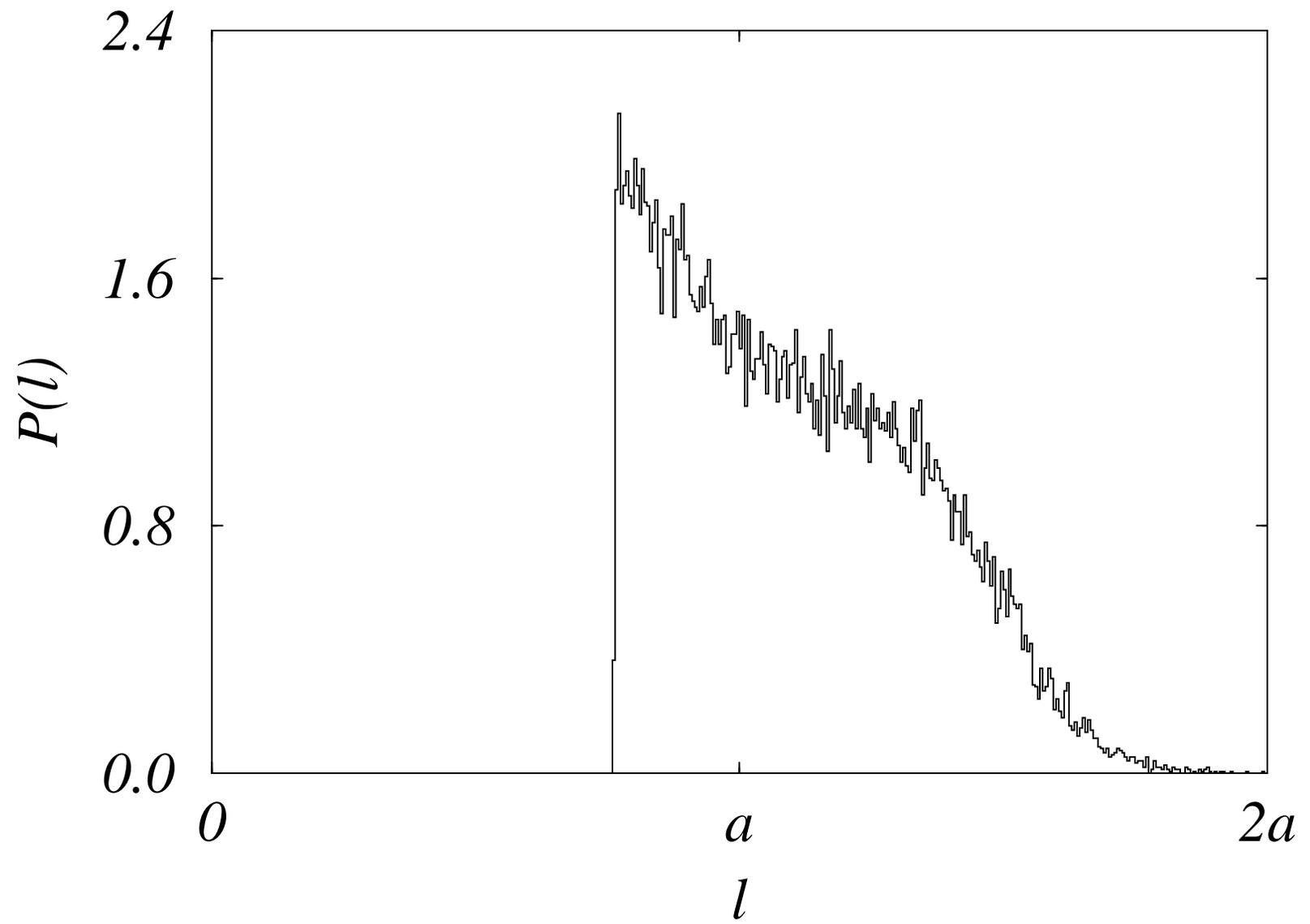
\item[Figure 8.] Distribution probability of link lengths on a random
lattice calculated with $\kappa=1.3 \;$.
The $a$ in abscisses stands for one lattice spacing.
The curve is normalized to 1.

\vskip 3mm

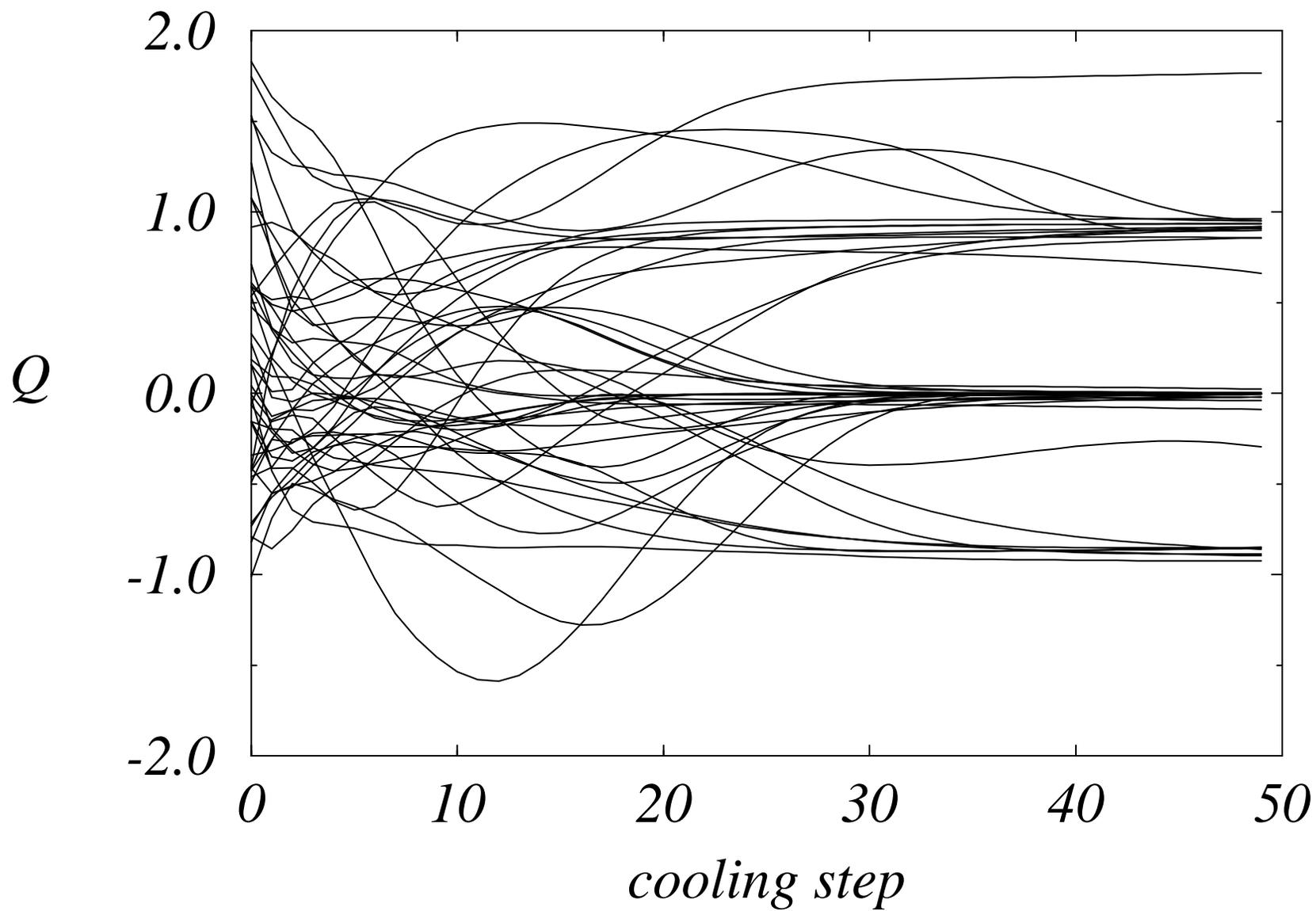
\item[Figure 9.] Evolution of the measured topological charge along
with 50 cooling steps for 40 uncorrelated configurations. Notice the
clustering towards integer values after a few coolings. A random
lattice with $40^2$ sites and $\kappa=\infty$ was
used at $\beta=1.4 \; $.

\vskip 3mm

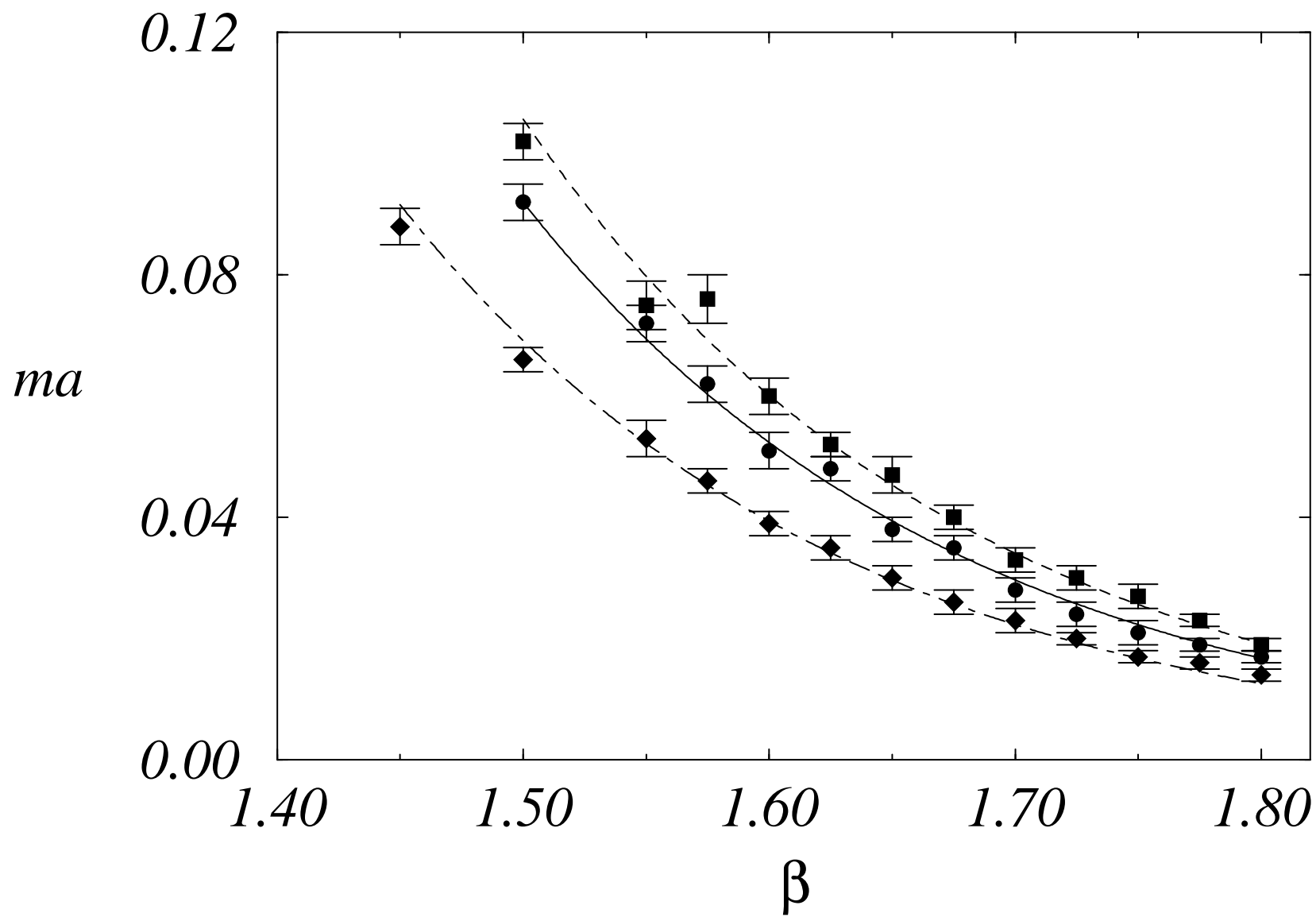
\item[Figure 10.] Monte Carlo data for the mass gap
on a lattice with $200^2$ sites. The lines are the results of the fits
performed on the Monte Carlo data (shown with circles, squares and
diamonds). The solid line (circles), dashed line
(squares) and dot dashed line (diamonds) correspond to the regular
lattice, $\kappa=1.3$ random lattice and $\kappa=\infty$ random
lattice respectively.

\vskip 3mm

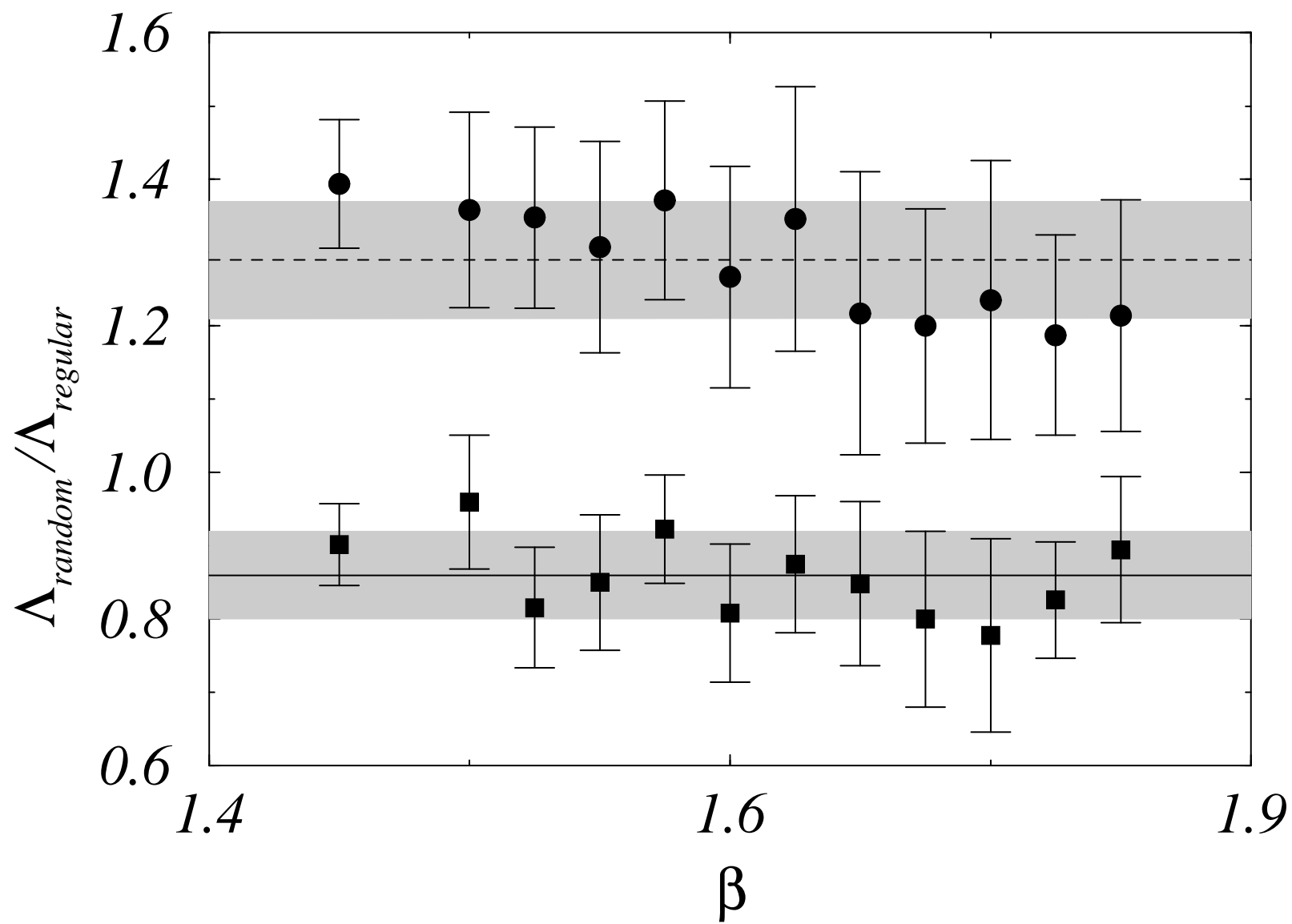
\item[Figure 11.] Ratio of $\Lambda$ parameters as obtained from the
Monte Carlo data of the mass gap. The solid line
(squares) and dashed line (circles) correspond to $\kappa=1.3$ and
$\kappa=\infty$ respectively. The average ratios are
$\Lambda_{\rm random}/\Lambda_{\rm regular}(\kappa=1.3) = 0.86(6)$ and
$\Lambda_{\rm random}/\Lambda_{\rm regular}(\kappa=\infty) = 1.29(8)$.

\vskip 3mm

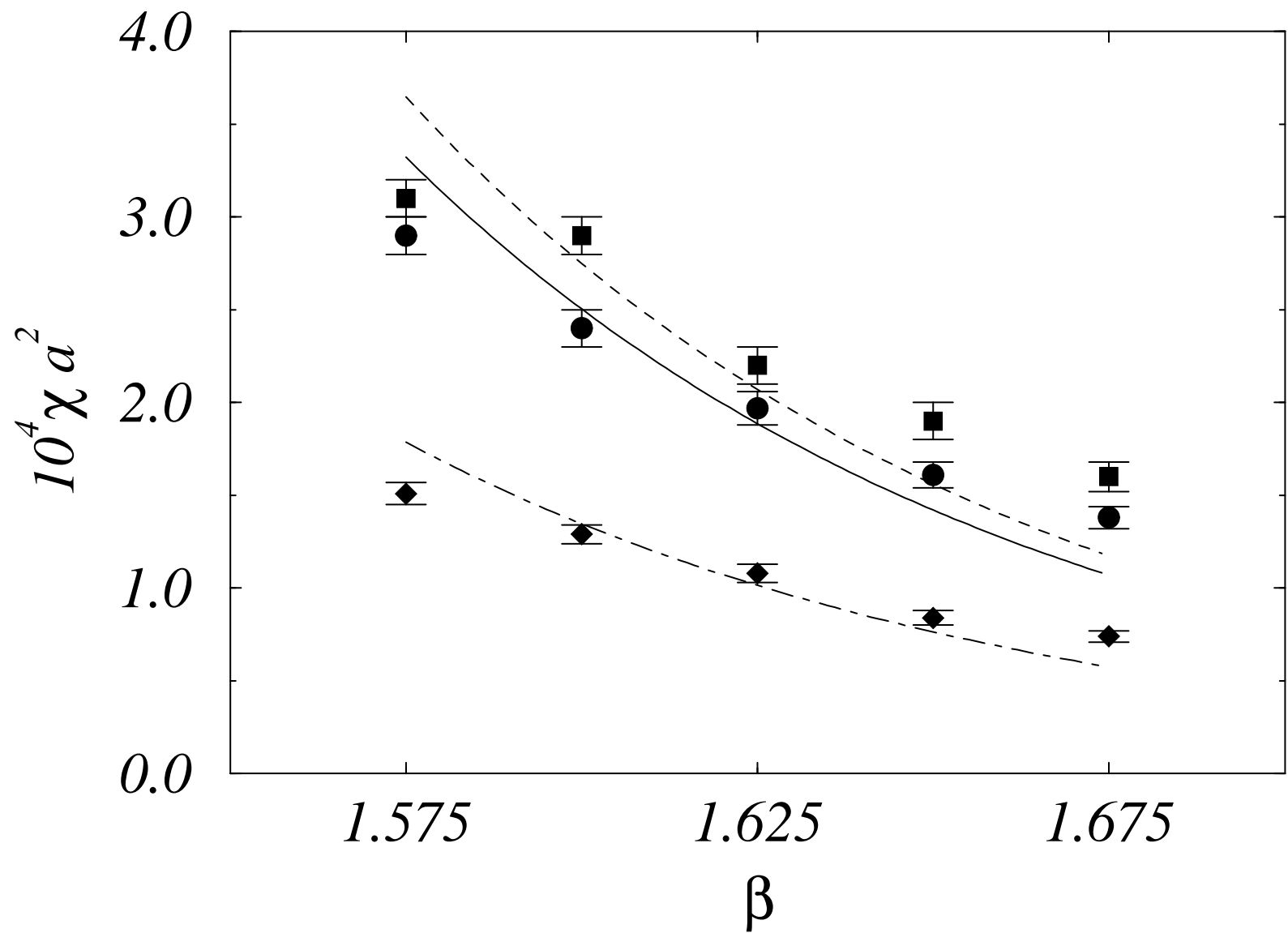
\item[Figure 12.] Monte Carlo data for the topological susceptibility
on a lattice with $200^2$ sites. The lines are the results of the fits
performed on the Monte Carlo data (shown with circles, squares and
diamonds). The solid line (circles), dashed line
(squares) and dot dashed line (diamonds) correspond to the regular
lattice, $\kappa=1.3$ random lattice and $\kappa=\infty$ random
lattice respectively.

\vskip 3mm

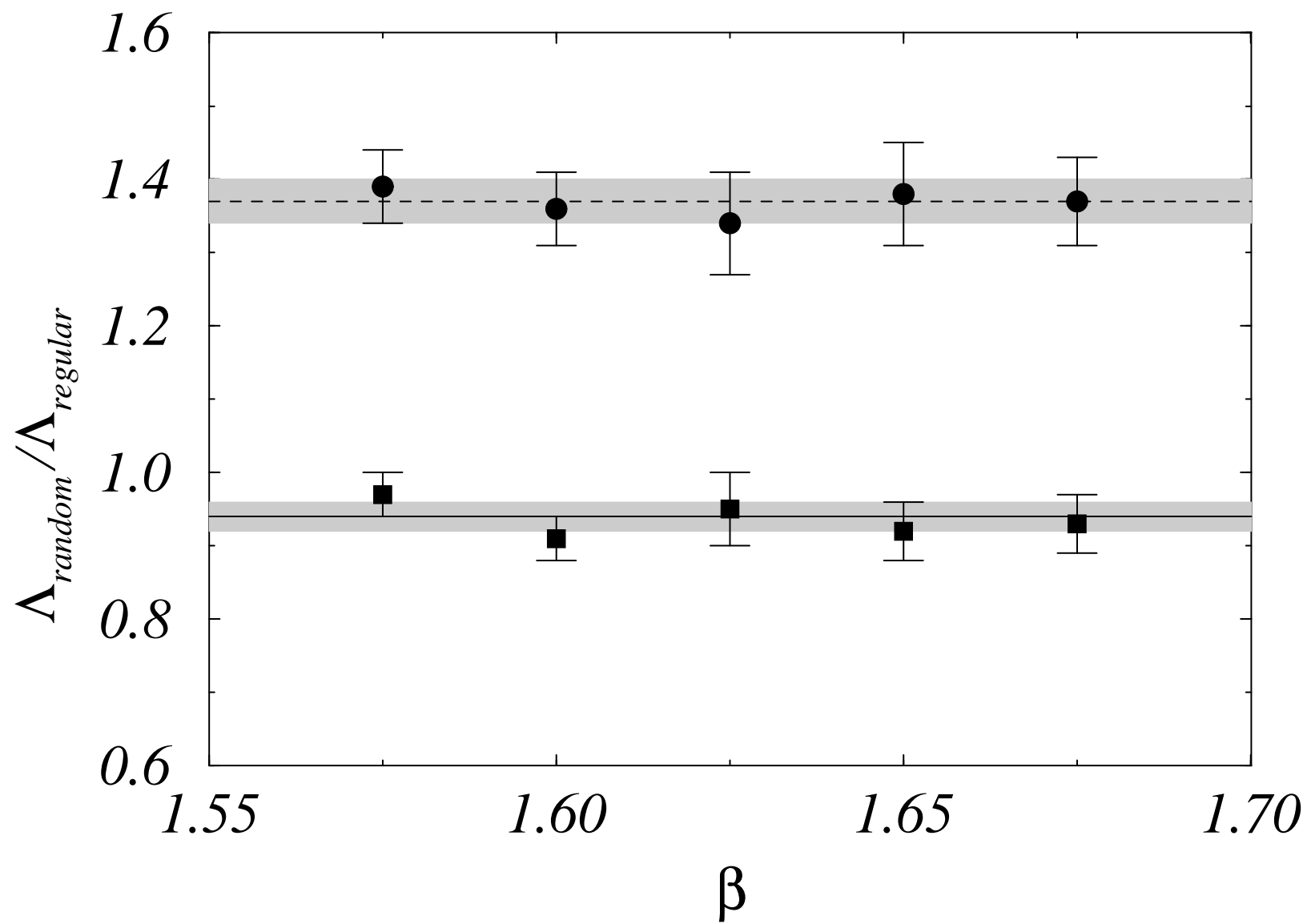
\item[Figure 13.] Ratio of $\Lambda$ parameters as obtained from the
Monte Carlo data of the topological susceptibility. The solid line
(squares) and dashed line (circles) correspond to $\kappa=1.3$ and
$\kappa=\infty$ respectively. The average ratios are
$\Lambda_{\rm random}/\Lambda_{\rm regular}(\kappa=1.3) = 0.94(2)$ and
$\Lambda_{\rm random}/\Lambda_{\rm regular}(\kappa=\infty) = 1.37(3)$.

\end{enumerate}

\vskip 2cm

\noindent{\bf Table captions}

\vskip 5mm

\begin{enumerate}

\item[Table 1.] Average values for the distance between a site and its
closest neighbour, $\left\langle r \right\rangle$ and the link length
$\left\langle l \right\rangle$ in units of lattice spacing
$a$ for several values of $\kappa$. These numbers have been
calculated numerically.

\vskip 3mm

\item[Table 2.] Monte Carlo data for the mass gap, $ma^2$ for regular
and random lattices with $200^2$ sites for the $O(3)$ $\sigma$ model.

\vskip 3mm

\item[Table 3.] Monte Carlo data for the mass gap, $ma^2$ for regular
and random lattices with $400^2$ sites for the $O(3)$ $\sigma$ model.

\vskip 3mm

\item[Table 4.] Monte Carlo data for the mass gap, $ma^2$ for regular
and random lattices with $400^2$ sites for the $O(4)$ $\sigma$ model.

\vskip 3mm

\item[Table 5.] Monte Carlo data for the mass gap, $ma^2$ for regular
and random lattices with $300^2$ sites for the $O(8)$ $\sigma$ model.

\vskip 3mm

\item[Table 6.] Monte Carlo data for the topological susceptibility,
$10^4 a^2 \chi$ for regular and random lattices with
$200^2$ sites for the
$O(3)$ $\sigma$ model.

\vskip 3mm

\item[Table 7.] Monte Carlo data for the topological susceptibility,
$10^4 a^2 \chi$ for a regular lattice with $300^2$ sites for the
$O(3)$ $\sigma$ model.

\vskip 3mm

\item[Table 8.] Monte Carlo data for the topological susceptibility,
$10^4 a^2 \chi$ for a $\kappa=\infty$ random lattice with $400^2$
sites for the $O(3)$ $\sigma$ model.

\end{enumerate}

\newpage

\centerline{\bf Table 1}

\vskip 1cm

$$\vbox{\tabskip=0pt  \offinterlineskip
\halign to250pt{\strut#& \vrule#\tabskip=1em plus2em& \hfil#& \vrule#&
\hfil#\hfil& \vrule#& \hfil#& \vrule#\tabskip=0pt\cr \noalign{\hrule}
& & \omit\hidewidth  $\kappa$\hidewidth& & \omit\hidewidth
  $\left\langle l \right\rangle \! /a$\hidewidth& & \omit\hidewidth
  $\left\langle r \right\rangle \! /a$\hidewidth& \cr \noalign{\hrule}
& & $\infty$& & 1.131(2)& & 0.508(2)& \cr \noalign{\hrule}
& & 1.3& & 1.113(1)& & 0.857(1)& \cr \noalign{\hrule}
& & 1.2& & 1.109(1)& & 0.90(2)& \cr \noalign{\hrule}
\noalign{\smallskip} }} $$

\vskip 2cm

\centerline{\bf Table 2}

\vskip 1cm

\moveright 1.3 in
\vbox{\offinterlineskip
\halign{\strut
\vrule \hfil\quad $#$ \hfil \quad &
\vrule \hfil\quad $#$ \hfil \quad &
\vrule \hfil\quad $#$ \hfil \quad &
\vrule \hfil\quad $#$ \hfil \quad \vrule \cr
\noalign{\hrule}
\beta & {\rm regular} & \kappa=1.3 & \kappa=\infty  \cr
\noalign{\hrule}
1.45 & -  & - & 0.0884(27) \cr
\noalign{\hrule}
1.5 	& 0.0918(32)	& 0.1018(30) 	& 0.0656(25) \cr
\noalign{\hrule}
1.55	& 0.0716(23)	& 0.0750(29)	& 0.0527(31) \cr
\noalign{\hrule}
1.575	& 0.0616(30)	& 0.0759(23)	& 0.0462(25) \cr
\noalign{\hrule}
1.6	& 0.0505(25)	& 0.0595(22)	& 0.0390(20) \cr
\noalign{\hrule}
1.625	& 0.0480(12)	& 0.0522(20)	& 0.0346(21) \cr
\noalign{\hrule}
1.65	& 0.0379(18)	& 0.0471(27)	& 0.0301(16) \cr
\noalign{\hrule}
1.675	& 0.0350(8)	& 0.0397(30)	& 0.0260(17) \cr
\noalign{\hrule}
1.7	& 0.0284(19)	& 0.0333(9)	& 0.0228(15) \cr
\noalign{\hrule}
1.725	& 0.0243(16)	& 0.0298(9)	& 0.0200(7) \cr
\noalign{\hrule}
1.75	& 0.0213(14)	& 0.0266(7)	& 0.0171(14) \cr
\noalign{\hrule}
1.775	& 0.0191(6)	& 0.0227(7)	& 0.0155(13) \cr
\noalign{\hrule}
1.8	& 0.0175(12)	& 0.0190(6)	& 0.0140(6) \cr
\noalign{\hrule}
}}

\vskip 2cm

\newpage

\centerline{\bf Table 3}

\vskip 1cm

\moveright 1.3 in
\vbox{\offinterlineskip
\halign{\strut
\vrule \hfil\quad $#$ \hfil \quad &
\vrule \hfil\quad $#$ \hfil \quad &
\vrule \hfil\quad $#$ \hfil \quad &
\vrule \hfil\quad $#$ \hfil \quad \vrule \cr
\noalign{\hrule}
\beta & {\rm regular} & \kappa=1.3 & \kappa=\infty  \cr
\noalign{\hrule}
1.6	& 0.0598(17)	& - 	& 0.0438(20) \cr
\noalign{\hrule}
1.65	& 0.0393(13)	& -	& 0.0314(20) \cr
\noalign{\hrule}
1.7	& 0.0290(12)	& -	& 0.0231(10) \cr
\noalign{\hrule}
1.75	& 0.0215(9)	& 0.0252(11)	& 0.0170(9) \cr
\noalign{\hrule}
1.775	& 0.0184(7)	& 0.0228(9)	& 0.0137(4) \cr
\noalign{\hrule}
1.8	& 0.0153(6)	& 0.0191(10)	& 0.0123(8) \cr
\noalign{\hrule}
}}

\vskip 2cm

\centerline{\bf Table 4}

\vskip 1cm

\moveright 1.3 in
\vbox{\offinterlineskip
\halign{\strut
\vrule \hfil\quad $#$ \hfil \quad &
\vrule \hfil\quad $#$ \hfil \quad &
\vrule \hfil\quad $#$ \hfil \quad &
\vrule \hfil\quad $#$ \hfil \quad \vrule \cr
\noalign{\hrule}
\beta & {\rm regular} & \kappa=1.3 & \kappa=\infty  \cr
\noalign{\hrule}
2.4	& 0.0386(14)	& 0.0426(33)	& 0.0333(12) \cr
\noalign{\hrule}
2.5	& 0.0313(17)	& 0.0320(22)	& 0.0248(11) \cr
\noalign{\hrule}
2.6	& 0.0211(9)	& 0.0228(16)	& 0.0179(8) \cr
\noalign{\hrule}
2.7	& 0.0165(7)	& 0.0165(12)	& 0.0129(6) \cr
\noalign{\hrule}
2.8	& 0.0117(4)	& 0.0126(9)	& 0.0099(5) \cr
\noalign{\hrule}
}}

\vskip 2cm

\centerline{\bf Table 5}

\vskip 1cm

\moveright 1.3 in
\vbox{\offinterlineskip
\halign{\strut
\vrule \hfil\quad $#$ \hfil \quad &
\vrule \hfil\quad $#$ \hfil \quad &
\vrule \hfil\quad $#$ \hfil \quad &
\vrule \hfil\quad $#$ \hfil \quad \vrule \cr
\noalign{\hrule}
\beta & {\rm regular} & \kappa=1.3 & \kappa=\infty  \cr
\noalign{\hrule}
4.6	& 0.1549(163)	& 0.1663(217)	& 0.0872(26) \cr
\noalign{\hrule}
5.2	& 0.0592(36)	& 0.0658(21)	& 0.0471(25) \cr
\noalign{\hrule}
5.8	& 0.0287(16)	& 0.0320(15)	& 0.0247(14) \cr
\noalign{\hrule}
}}

\vskip 2cm

\newpage

\centerline{\bf Table 6}

\vskip 1cm

\moveright 1.3 in
\vbox{\offinterlineskip
\halign{\strut
\vrule \hfil\quad $#$ \hfil \quad &
\vrule \hfil\quad $#$ \hfil \quad &
\vrule \hfil\quad $#$ \hfil \quad &
\vrule \hfil\quad $#$ \hfil \quad \vrule \cr
\noalign{\hrule}
\beta & {\rm regular} & \kappa=1.3 & \kappa=\infty  \cr
\noalign{\hrule}
1.575	& 2.90(13)	& 3.13(14)	& 1.511(65) \cr
\noalign{\hrule}
1.6	& 2.45(10)	& 2.85(13)	& 1.252(53) \cr
\noalign{\hrule}
1.625	& 1.97(9)	& 2.16(10)	& 1.078(51) \cr
\noalign{\hrule}
1.65	& 1.61(8)	& 1.89(10)	& 0.838(37) \cr
\noalign{\hrule}
1.675	& 1.38(6)	& 1.60(8)	& 0.741(33) \cr
\noalign{\hrule}
}}

\vskip 2cm

\centerline{\bf Table 7}

\vskip 1cm

\moveright 2.4 in
\vbox{\offinterlineskip
\halign{\strut
\vrule \hfil\quad $#$ \hfil \quad &
\vrule \hfil\quad $#$ \hfil \quad \vrule \cr
\noalign{\hrule}
\beta & {\rm regular} \cr
\noalign{\hrule}
1.575 & 2.99(13) \cr
\noalign{\hrule}
1.6 & 2.50(11) \cr
\noalign{\hrule}
1.625 & 1.96(9) \cr
\noalign{\hrule}
1.65 & 1.51(7) \cr
\noalign{\hrule}
1.675 & 1.37(6) \cr
\noalign{\hrule}
}}

\vskip 2cm

\centerline{\bf Table 8}

\vskip 1cm

\moveright 2.4 in
\vbox{\offinterlineskip
\halign{\strut
\vrule \hfil\quad $#$ \hfil \quad &
\vrule \hfil\quad $#$ \hfil \quad \vrule \cr
\noalign{\hrule}
\beta & \kappa=\infty \cr
\noalign{\hrule}
1.575 & 1.515(66) \cr
\noalign{\hrule}
1.6 & 1.307(79) \cr
\noalign{\hrule}
1.625 & 1.159(96) \cr
\noalign{\hrule}
1.65 & 0.939(41) \cr
\noalign{\hrule}
1.675 & 0.747(35) \cr
\noalign{\hrule}
}}

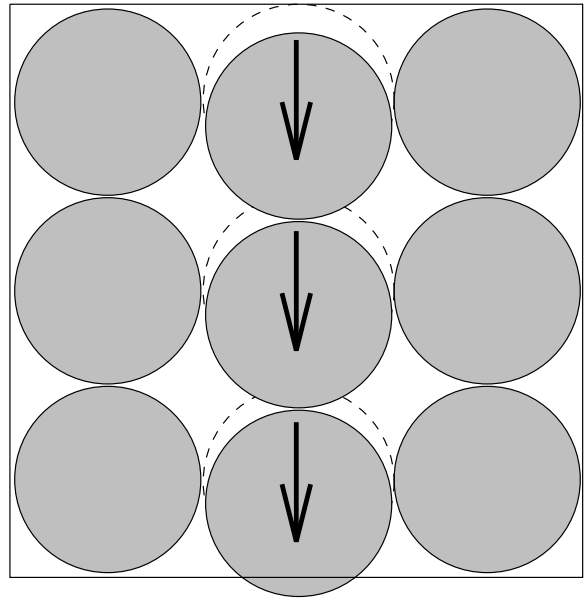
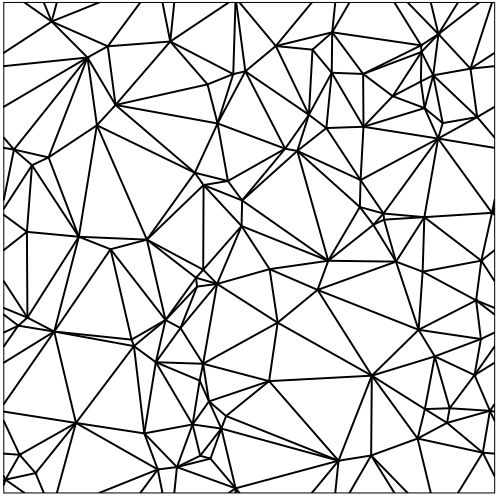
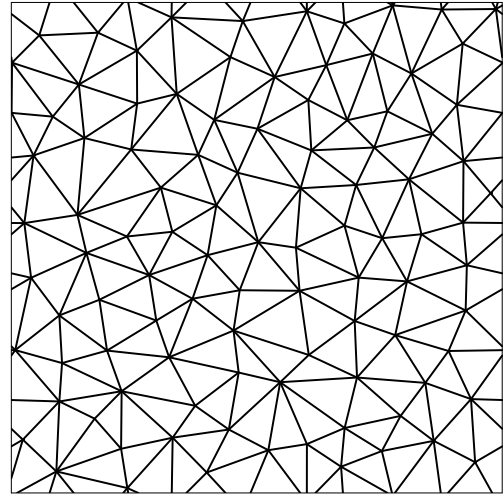
\end{document}